\documentclass[a4paper,11pt]{article}
\usepackage{pos}
\usepackage{
    amsmath,			
    amsthm,
    amsfonts,
    babel,
    bbold,
    bm,
    braket,
    caption,
    comment,
    dsfont,
    enumerate,		    
    enumitem,
    epsfig,
    fancyhdr,
    float,
    graphics,
    graphicx,			
    lastpage,			
    multicol,			
    multirow,			
    natbib,
    slashed,
    subcaption,
    tabularx,
    tikz,
    url
}
\usepackage[labelfont=bf]{caption}
\usepackage[export]{adjustbox} 
\usepackage[utf8]{inputenc}  
\DeclareUnicodeCharacter{3164}{ }

\usepackage{tikz}
\usetikzlibrary{external}
\usepackage{tikz-3dplot}
\usepackage{ifthen}
\usetikzlibrary{shapes.geometric, arrows}
\usetikzlibrary{arrows.meta}
\tikzstyle{startstop} = [rectangle, rounded corners, minimum width=3cm, minimum height=1cm,text centered, draw=black, fill=red!30]
\tikzstyle{process} = [rectangle, minimum width=3cm, minimum height=1cm, text centered, draw=black, fill=orange!30]
\tikzstyle{decision} = [diamond, minimum width=3cm, minimum height=1cm, text centered, draw=black, fill=green!30]
\tikzstyle{answer} = [rectangle, rounded corners, minimum width=3cm, minimum height=1cm,text centered, draw=black, fill=blue!30]
\tikzstyle{arrow} = [thick,->,>=stealth]
\tikzset{Stealth/.style={Stealth[scale=1]}}
\definecolor{fillcolorblue}{RGB}{122,203,251} 
\definecolor{bordercolorblue}{RGB}{46,132,253} 
\definecolor{fillcolorpurple}{RGB}{171,141,246} 
\definecolor{bordercolorpurple}{RGB}{103,38,173} 
\definecolor{fillcolororange}{RGB}{246,166,119} 
\definecolor{bordercolororange}{RGB}{245,104,48} 

\title{A Variational Approach to Quantum Field Theory}

\author*[a]{M. Rovira}
\author[b]{ A. Parreño}
\author[b]{R.J. Perry}

\usepackage{soul}

\affiliation[a]{Facultat de Física, Universitat de Barcelona,\\ C/ Martí i Franquès 1-11, 08028 Barcelona, Spain \\ ㅤ}
\affiliation[b]{Dept. de Física Quàntica i Astrofísica \& Institut de Ciències del Cosmos, Universitat de Barcelona,\\ C/ Martí i Franquès 1-11, 08028 Barcelona, Spain \\ ㅤ}

\emailAdd{martirovira2000@gmail.com}
\emailAdd{assum@fqa.ub.edu}
\emailAdd{perryrobertjames@gmail.com}

\abstract{
In strongly coupled field theories, perturbation theory cannot be employed to study the low-energy spectrum. 
Thus, non-perturbative techniques are required. We employ the variational method, a rigorous, non-perturbative approach which provides variational upper bounds on the energy eigenstates of the theory. An essential step in the variational method is the choice of trial wave function. In this work, we study the viability of employing a neural network as our variational ansatz. As a first step towards phenomenologically interesting strongly coupled theories like quantum chromodynamics, we study scalar field theories with quartic couplings.
}

\FullConference{%
  QNP2024 - The 10th International Conference on Quarks and Nuclear Physics\\
  8-12 July 2024\\
  Avinguda Diagonal 643, 08028 Barcelona, Spain
}

\begin{document}
\maketitle
\section{Introduction}

Given a quantum mechanical Hamiltonian, $\hat{H}$, a general, non-perturbative approach to studying the energy eigenvalues is the variational method. This approach is based on the observation that for any normalizable trial wave function $\ket{\Psi}$, the expectation value of the Hamiltonian provides an upper bound on the ground state energy, $E_0$ \cite{principles-qm}.
The variational method proceeds by first postulating a flexible \textit{ansatz} for the wave function which depends on a number of free parameters, which may be varied to lower the expectation value of the Hamiltonian. The central challenge to variational calculations is the selection of a suitable \textit{ansatz}. 

Neural networks (NN) are a computational model employed in machine learning. For a variety of NN activation functions, it has been shown that they satisfy a \textit{Universal Approximation Theorem}, which guarantees that with a suitably complex NN, it is possible to approximate a function, $f:\mathrm{R}^n\to\mathrm{R}$ such that $|f-\text{NN}|<\epsilon$ for any $\epsilon>0$. In addition, NN software packages implement auto-differentiation which enables the automatic computation of gradients required for parameter optimization. Therefore, it is interesting to explore the possibility of employing a NN as a variational \textit{ansatz}. Here, we study simple quantum mechanical systems using the variational method, with a NN trial wave function. In particular, in this proceedings, we describe the determination of a ground state variational bound for interacting scalar field theories in $d=0+1$ and $d=3+1$ dimensions.


\section{Methods}
\subsection{Lattice Hamiltonians}
The Hamiltonian $\hat{H}$ for an anharmonic oscillator is given by
\begin{equation} \label{eq:hamiltonian-qm}
    \hat{H}=\frac{\hat{p}^2}{2m}+\frac{1}{2}\mu^2\hat{x}^2 +\lambda \hat{x}^4\,,
\end{equation}
where $\hat{p}$ and $\hat{x}$ are the momentum and position operators, respectively, $m$ is the mass and $\mu$ is the stiffness. In the non-interacting system where $\lambda = 0$, it can be shown that the normalized energy eigenfunctions are
\begin{equation}
    \braket{x|\psi_n}=\psi_n(x)=\frac{1}{\sqrt{2^n \hspace{1mm}n!}}\left( \frac{\sqrt{m}\mu}{\pi}\right)^{1/4}e^{-\frac{\sqrt{m}\mu}{2}x^2}H_n\left(\left[ \sqrt{m} \mu x^2\right]^{1/2}\right)\,,
\end{equation}
where $n=0,1,2\dots$, $H_n(x)$ are the Hermite polynomials and $E_n=(1/2+n)\mu/\sqrt{m}$ are the energy eigenvalues. The ground state energy eigenfunction is $\psi_0(x) \propto \exp\left(-\sqrt{m}\mu x^2/2\right)$.

In $d=3+1$, the continuum $\Phi^4$ theory is defined by the Hamiltonian
\begin{equation}
    \begin{split}
        \hat{H} = \int d^3x \left(\frac{1}{2}\hat{\Pi}^2(\bm{x})+\frac{1}{2}(\nabla\hat{\Phi}(\bm{x}))^2+\frac{1}{2}m^2\hat{\Phi}^2(\bm{x})+\lambda\hat{\Phi}^4(\bm{x})\right)\,,
    \end{split}
\end{equation}
where $\hat{\Phi}$ and $\hat{\Pi}$ are the field operator and its conjugate momentum. The lattice discretized version of this Hamiltonian in position representation with box length $L$ and periodic boundary conditions is defined in Ref.~\cite{Gattringer2010} as
\begin{equation}\label{eq:hamiltonian-lattice}
    \begin{split}
        &H  =  \sum_{\bm{n}\in\Lambda}\left( -\frac{1}{2}\frac{\partial^2}{\partial \Phi(\bm{n})^2}+ \frac{1}{2}\sum_{j=1}^3\left(\frac{\Phi(\bm{n}+\hat{j})-\Phi(\bm{n}-\hat{j})}{2}\right)^2+\frac{1}{2}m^2\Phi(\bm{n})^2+\lambda\Phi^4(\bm{n})\right)\,,
    \end{split}
\end{equation}
where the lattice sites are $\Lambda = \{\mathbf{n}|n_i=1,2,\dots,L\text{ for }i=1,2,3 \}$, the lattice spacing has been set to unity, $\hat{j}$ are unit vectors and $m$ is the mass of the field. One can show that the ground state wavefunctional for the non-interacting system is given by
\begin{equation}
    \psi_0(\Phi(\mathbf{n}_1),\dots,\Phi(\mathbf{n}_{L^3})) = \frac{1}{N}\exp \left(-\frac{1}{2}\sum_{i,j}\Phi(\bm{n}_i) K_{ij}\Phi(\bm{n}_j)\right)\,,
\end{equation}
where $N$ is a normalization constant and $K$is the kernel defined as $K^2_{ij}=\mu^2 \delta_{ij}-\frac{1}{4}\left(\delta_{i+1,j}-\delta_{i-1,j}\right)^2$.

\subsection{The Variational Method}
We use the variational method to study the energy eigenvalues of the Hamiltonians described in the previous section. We employ a trial wave function $\ket{\Psi;\alpha}$ which depends on the variational parameters $\alpha =\{ \alpha_0, \alpha_1, \dots \}$ and optimize these parameters by minimizing the energy of the system given by
\begin{equation}
    \braket{H}(\alpha)=\int D\sigma\, \Psi^*(\sigma;\alpha) H \Psi(\sigma;\alpha) = \int D\sigma |\Psi(\sigma;\alpha)|^2 \Psi^{-1}(\sigma;\alpha)  H \Psi(\sigma;\alpha)\,,
\end{equation}
where $\sigma$ are the position-space generalized coordinates of the specific system. The measure for the two systems studied is
\begin{equation}
D\sigma=
\begin{cases}
\hphantom{\prod_{\mathbf{n}\in\Lambda}}\int_{-\infty}^\infty dx\,,
\\
\prod_{\mathbf{n}\in\Lambda} \int_{-\infty}^\infty d\Phi(\mathbf{n})\,,
\end{cases}
\end{equation}
respectively. By defining the local energy $E_{\text{loc}}(\sigma;\alpha) = \Psi^{-1}(\sigma;\alpha)H\Psi(\sigma;\alpha)$, the expectation value of the Hamiltonian can be computed using Monte Carlo techniques. Specifically, the Monte Carlo estimate for the energy is
\begin{equation}
    \braket{H}(\alpha) \approx \frac{1}{\text{N}_{\text{cf}}}\sum_{i} E_{\text{loc}}(\sigma^{(i)};\alpha)\,,
\end{equation}
where $\sigma^{(i)}$ with $i=1...,\text{N}_{\text{cf}}$ are field configurations distributed according to $|\Psi(\sigma;\alpha)|^2$. In practice, we use the Metropolis-Hastings algorithm to generate these configurations~\cite{Metropolis:1953am,Hastings:1970aa}. 

\begin{figure}
    \centering
    \begin{subfigure}{0.4\textwidth}
        \centering
        \includegraphics[width=\textwidth]{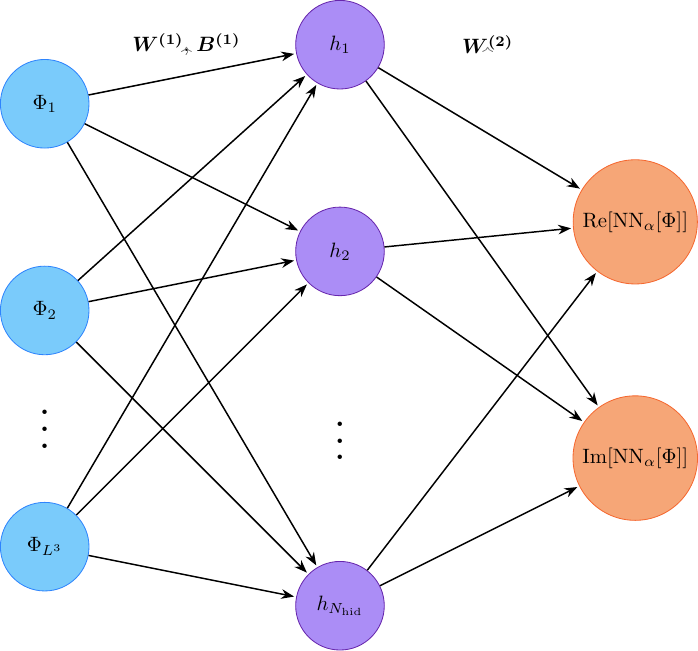}
    \end{subfigure}\hfill
    \begin{subfigure}{0.5\textwidth}
        \centering
        \includegraphics[width=\textwidth]{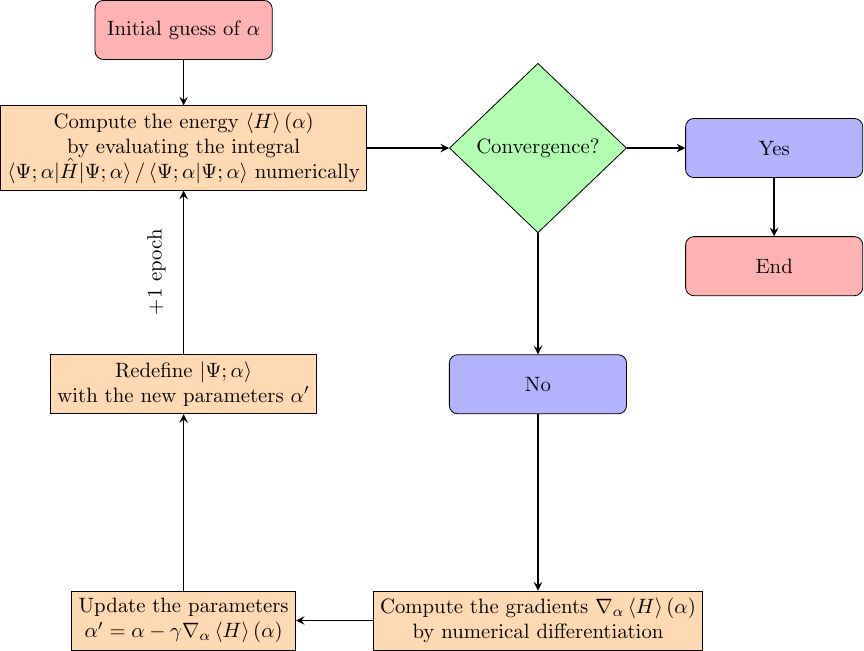}
    \end{subfigure}
    \caption{Left: Neural network architecture serving as the ansatz for a quantum field system in a $L^3$ lattice. Right: Flowchart of the VMC algorithm.}
    \label{fig:NN-VMC}
\end{figure}

For this study, we use 
\begin{equation}
    \Psi(\sigma;\alpha) = \psi_0(\sigma)\cdot \text{NN}(\sigma;\alpha)\,,
\end{equation}
for both the anharmonic oscillator and $\Phi^4$ systems. $\text{NN}(\sigma;\alpha)$ is a neural network whose inputs are the values of the generalized coordinates of the system and $h_i$ are the activation functions, which in this work are chosen to be sigmoids. The weights ($\bm{W^{(1),(2)}}$) and the biases ($\bm{B^{(1)}}$) of the NN are the variational parameters. In the case of a field theory, the NN architecture is visualized in Fig.~\ref{fig:NN-VMC}. We employ the Variational Monte Carlo (VMC) algorithm~\cite{Battaglia2023}, also pictured in Fig.~\ref{fig:NN-VMC}, to obtain variational estimates of the ground state energy. 

\section{Results}
\begin{figure}
    \centering
    \begin{subfigure}{0.465\textwidth}
        \centering
        \includegraphics[width=\linewidth]{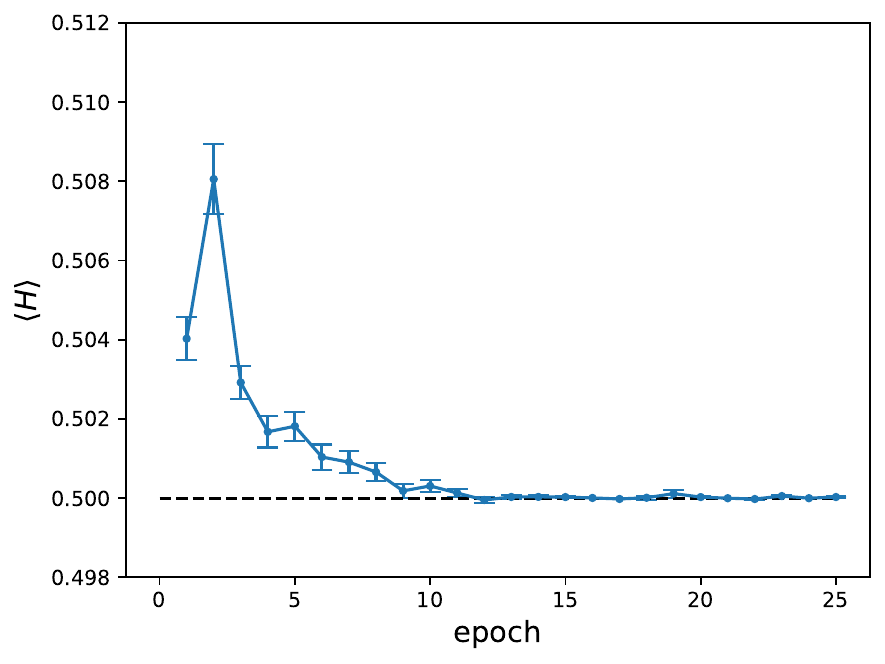} 
    \end{subfigure}\hfill
    \begin{subfigure}{0.45\textwidth}
        \centering
        \includegraphics[width=\linewidth]{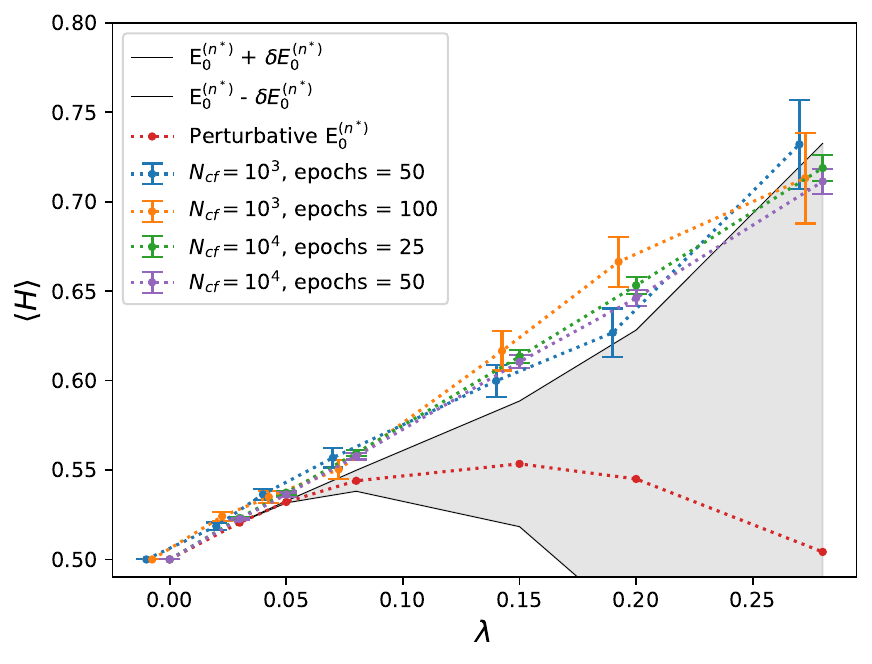} 
    \end{subfigure}
    \begin{subfigure}{0.465\textwidth}
        \centering
        \includegraphics[width=\linewidth]{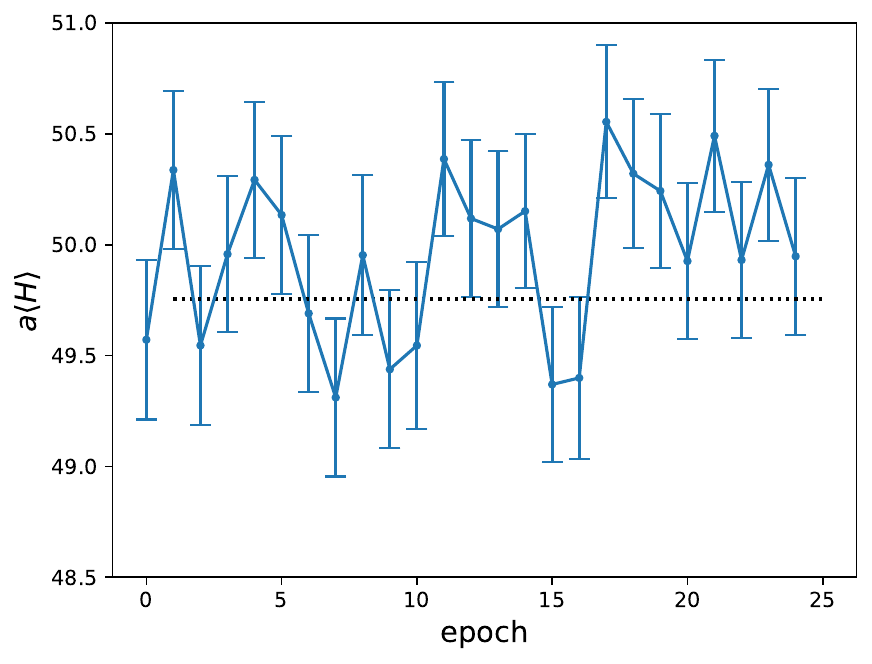} 
    \end{subfigure}\hfill
    \begin{subfigure}{0.45\textwidth}
        \centering
        \includegraphics[width=\linewidth]{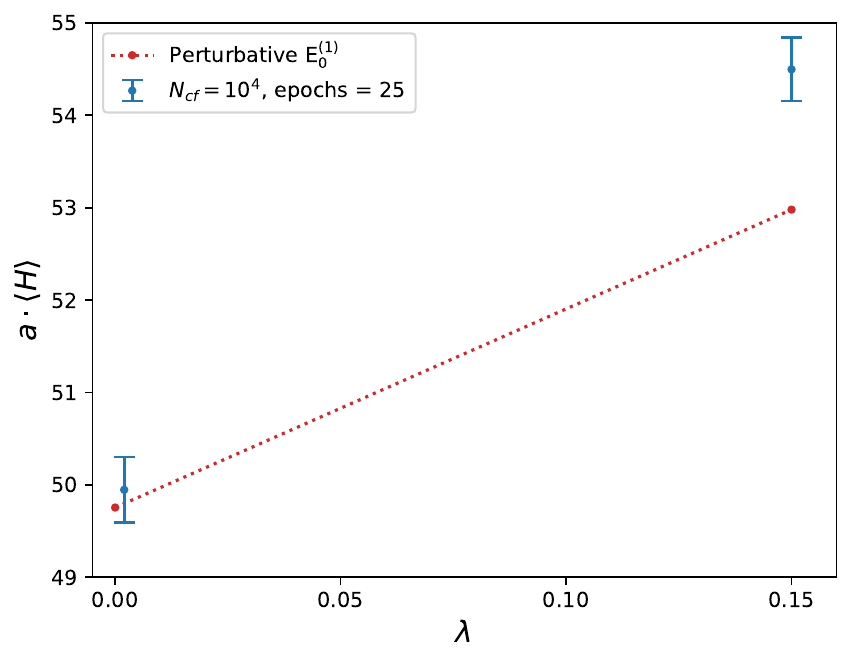} 
    \end{subfigure}
    \caption{
    Left: Variational bound on the ground state energy for the anharmonic oscillator (top) and $\Phi^4$ field theory (bottom), obtained from the gradient descent algorithm as a function of the epoch for N$_{\text{cf}}=10^4$, where the exact ground state energy is also plotted (black dashed line).  Right: Variational bound’s dependence on the strength of the coupling for the anharmonic oscillator (top) and $\Phi^4$ field theory (bottom). The shaded band shows the perturbative result taken from Ref.~\cite{Bender}.}
    \label{fig:va-bounds}
\end{figure}

In Fig.~\ref{fig:va-bounds} (top-left) the energy evolution for a harmonic oscillator ($\lambda=0$) with $\mu = m = 1$ is shown and compared with its exact value. It shows approximate saturation of the bounds (i.e. $\braket{H}(\alpha)\approx E_0$) for this system and after $\sim$ 10 epochs we already achieve a good approximation of the ground state. Results for different $\lambda$, number of epochs and sizes of Monte Carlo samplings are portrayed in Fig.~\ref{fig:va-bounds} (top-right) together with the perturbative values from Ref. \cite{Bender}. As these perturbative values are obtained by truncating an asymptotic series, they present an uncertainty. We have estimated its value by
\begin{equation}\label{CH2-RESULTS-AHO-error-pert}
    \delta E_0^{(n^*)} = \frac{1}{2}\left(E_0^{(n^*+1)}-E_0^{(n^*)}\right)\,,
\end{equation}
where $n^*$ is the order at which the series truncated. In the plot, it can be observed that the variational results fall always, within a $1\sigma-2\sigma$ margin of error, inside the perturbative band generated by this error. 

Having obtained satisfying results and performance of these simpler systems, the same operations have been applied to the $\Phi^4$ field theories in an $L=4$ cubic lattice, showing also satisfactory convergence and bound saturation in Fig.~\ref{fig:va-bounds} (bottom-left), as well as agreement between the variational and perturbative results in Fig.~\ref{fig:va-bounds} (bottom-right). The first contrast is that the quantum mechanical system, takes 42 s/epoch while the field system takes 1h 20 min/epoch, meaning that the latter is 115 times more efficient. This is due to the fact that the algorithm performance scales with the lattice volume.

\section{Summary and Future Outlook}
Using the neural network ansatz, we obtained variational upper bounds in agreement with the perturbative results for scalar field theories in $d=0+1$ and $d=3+1$ dimensions and different values of the coupling. The algorithm takes around 10-15 epochs to achieve convergence. For the $d=3+1$ system, less values of $\lambda$ are explored, due to the computational cost of this system. Possible future research directions could focus on optimizing the algorithm so that the computational cost is lower allowing further exploration of $\lambda$, computing eigenvalues above the ground state, and studying other quantum field theories with this approach.

\bibliographystyle{JHEP}
{\bibliography{refs}}

\end{document}